%% file: main.tex
\documentclass[conference]{IEEE-conference-template-062824/IEEEtran}

\usepackage{subcaption}
\usepackage{tabularx}
\usepackage{booktabs}
\usepackage{graphicx}
\usepackage{amsmath}
\usepackage{comment}

\begin{document}

\title{HORIZON: A Read-Efficient Firmware for DNA Storage with Horizontal Layout}

%\author{\IEEEauthorblockN{Anonymous Authors}}

\author{
  \IEEEauthorblockN{
    Alex Sensintaffar\IEEEauthorrefmark{3},
    Roop Kiran\IEEEauthorrefmark{2},
    Yang Chen\IEEEauthorrefmark{3},
    Mai Zheng\IEEEauthorrefmark{2}\IEEEauthorrefmark{1},
    Bingzhe Li\IEEEauthorrefmark{3}\IEEEauthorrefmark{1}
  }
  
  \IEEEauthorblockA{\IEEEauthorrefmark{3}University of Texas at Dallas, \IEEEauthorrefmark{2}Iowa State University}
  \IEEEauthorblockA{\IEEEauthorrefmark{1}Corresponding author}
}

\maketitle

\begin{abstract}
DNA storage is a promising medium for long-term archiving, but its read performance is limited by coarse-grained random access. Existing random-access DNA storage designs suffer from high read amplification because their sequential layouts co-locate frequently and infrequently accessed data under the same primer pair, where any read must retrieve all associated strands even when only a small fraction is needed. We present HORIZON, a read-efficient allocation policy for DNA block devices that reduces read amplification through activity-aware horizontal placement. HORIZON first introduces a horizontal layout distributing writes round-robin across primer pairs, rather than filling each sequentially. It classifies newly written blocks in the write buffer as active or inactive, tracks recent primer-pair accesses using a sliding-window temperature model, and allocates blocks based on block activity and primer-pair occupancy. Simulations show HORIZON consistently reduces read amplification compared with state-of-the-art schemes across MSR and FIU traces and synthetic filesystem workloads.

\end{abstract}

%% Note: Classification and Keywords are only required for the camera-ready version

% <concept_id>10010520.10010575.10010577</concept_id>
% <concept_desc>Computer systems organization~Reliability</concept_desc>
% <concept_id>10010520.10010553.10010562</concept_id>
% <concept_desc>Computer systems organization~Embedded systems</concept_desc>
% <concept_significance>100</concept_significance>
% <concept_id>10010583.10010588.10010592</concept_id>
% <concept_desc>Hardware~External storage</concept_desc>
% <concept_significance>300</concept_significance>

% \ccsdesc[500]{Computer systems organization~Reliability}
% \ccsdesc[100]{Computer systems organization~Embedded systems}
% \ccsdesc[300]{Hardware~External storage}
% \ccsdesc[300]{Hardware~Sensor applications and deployments}
% \ccsdesc[300]{Hardware~Wireless integrated network sensors}

\sloppy
% Intro- Roop

\input{intro}

\input{background}

\input{motivation}

\section{Design}
\label{sec:design}

\begin{figure}[t]
    \centering
    \includegraphics[width=0.49\textwidth]{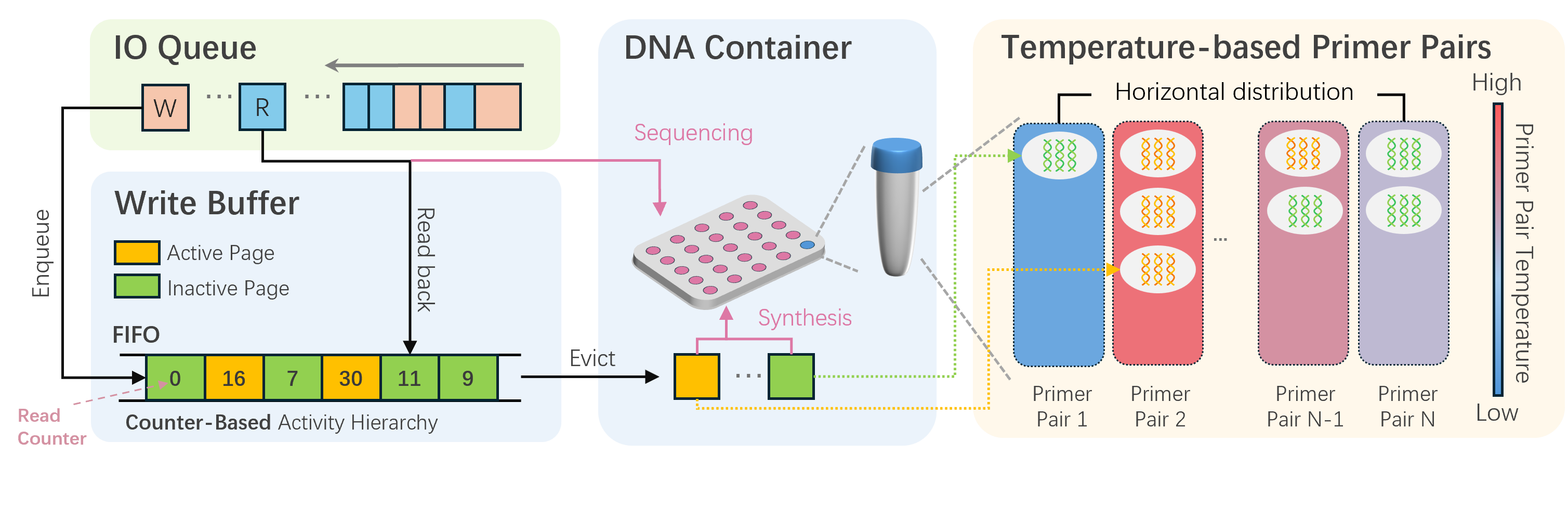}
    % \caption{The overall scheme of DNA Block Device.}
    \caption{HORIZON write-placement workflow.}
    \label{fig:overall_scheme}
\end{figure}

%This section presents HORIZON, a read-efficient firmware for DNA storage with horizontal layout for reducing read amplification in random-access DNA storage. The design builds on the observation that a primer pair acts as the minimum random-read unit for DNA retrieval. When a block is read, the system retrieves all DNA strands associated with the selected primer pair, even if only a small number of strands are needed for the requested block. As more data is written under a primer pair, the amount of unrelated DNA retrieved during each read increases. Therefore, read amplification grows as the storage system fills.
This section presents HORIZON, a read-efficient DNA storage firmware that uses horizontal layout to reduce read amplification in random-access DNA storage. HORIZON builds on the observation that each primer pair is the minimum random-read unit: reading one block retrieves all strands associated with its primer pair. As more data accumulates under a primer pair, each read retrieves more unrelated DNA, increasing read amplification.

HORIZON reduces this growth by changing how newly written data blocks are assigned to primer pairs. The design does not change the physical DNA medium, primer chemistry, strand encoding, or metadata mapping structure. Instead, it changes the data-placement policy used when buffered writes are committed to DNA. Figure~\ref{fig:overall_scheme} illustrates the overall workflow. Writes first enter the electronic write buffer, where they remain until they are flushed to DNA. While a block is buffered, reads to that block are served from the buffer and counted as evidence of early block activity. When the buffer reaches capacity, it flushes blocks. For each flushed block, HORIZON classifies the block as active or inactive based on its buffered read count, selects a primer pair using the corresponding placement rule, writes the block to DNA, and updates the logical-to-physical metadata so future reads locate the newest version of the block.

%HORIZON combines three placement signals. First, it distributes blocks horizontally across primer pairs so that occupancy grows more evenly instead of filling one primer pair before moving to the next. Second, it tracks a logical temperature for each primer pair using a sliding window of recent accesses. This temperature identifies recently active retrieval units while allowing short bursts to decay over time. Third, it uses the write buffer as an observation window for newly written blocks. Blocks whose buffered read count reaches the activity threshold are classified as active, while blocks whose buffered read count remains below the threshold are classified as inactive. At flush time, active blocks prioritize low-occupancy primer pairs to reduce immediate read amplification, while inactive blocks prefer colder primer pairs to avoid polluting recently active retrieval units.

%This design is well suited to DNA block devices because writes are slow, buffering is practical, and erase operations are expensive. Once data is written to DNA, it cannot be cheaply migrated or selectively removed; overwritten versions remain physically present until a larger erase operation destroys the containing DNA pool. HORIZON therefore focuses on making better initial placement decisions before data is committed to DNA. By combining horizontal placement, buffered activity classification, primer-pair temperature, and occupancy-aware placement, HORIZON reduces unnecessary growth in read amplification while leaving the underlying DNA storage process unchanged.

\subsection{Horizontal Primer-Pair Layout}
The first design choice in HORIZON is to distribute newly written blocks horizontally across primer pairs. Unlike sequential allocation, which fills one primer-pair group before moving to the next, horizontal placement spreads data across the primer library so that primer-pair occupancy grows more evenly. This layout reduces read amplification during early and intermediate device utilization because each primer-pair group contains fewer unrelated strands. When a block is read, the system still retrieves the entire selected primer-pair group, but that group is less crowded than it would be under vertical allocation.

Horizontal placement provides the foundation for the activity-aware policy described in the following subsections. A simple horizontal allocator, such as round-robin, can reduce read amplification by balancing occupancy, but it remains activity-oblivious. HORIZON therefore extends horizontal placement with primer-pair temperature and write-buffer activity classification, allowing active and inactive blocks to be placed differently while preserving the balancing benefit of horizontal layout.

\subsection{Primer-Pair Temperature}
After distributing data horizontally, HORIZON tracks which primer pairs have been accessed recently. We define the temperature of a primer pair as a logical measure of recent read and write activity. Temperature does not represent a physical property of the DNA medium; it is maintained by the firmware to guide future placement decisions.

For primer pair $p_i$, HORIZON computes temperature over a sliding window $W$ of recent operations:
\begin{equation}
T(p_i) = \#\{\text{reads and writes to } p_i \text{ within window } W\}.
\end{equation}

The window is defined by a fixed number of operations. When an operation enters the window, it increases the temperature of the primer pair it accesses. When that operation leaves the window, its contribution is removed. This decay prevents a primer pair from remaining permanently hot after a short burst of activity.

Primer-pair temperature helps HORIZON avoid placing inactive data into recently active retrieval units. A high temperature indicates recent access activity, so adding inactive data to that primer pair can increase unrelated DNA retrieval during future reads. HORIZON therefore uses temperature together with occupancy during placement: inactive blocks prefer colder primer pairs, while active blocks primarily prefer low-occupancy primer pairs and use temperature as a secondary signal.

% \subsection{Primer-Pair Temperature}

% The system assigns each primer pair a temperature that represents its recent access activity. Temperature is a logical property of the storage system; there is no physical difference between hot and cold primer pairs. A primer pair becomes hotter when recent reads or writes access data stored under that primer pair. These accesses are counted within a sliding window, which may be defined either by a fixed number of operations or by a fixed time interval.

% For primer pair $p_i$, the temperature is:

% \begin{equation}
% T(p_i) = \#\{\text{reads and writes to } p_i \text{ within window } W\}.
% \end{equation}

% When a new operation enters the window, it contributes to the temperature. When an operation leaves the window, it no longer contributes, allowing the temperature to decay over time. This prevents primer pairs from remaining permanently hot after a short burst of activity. A write updates temperature when the buffered block is flushed and physically written to DNA.

\subsection{Write Buffer Activity Classification}
\label{sec:write_buffer}
% DNA storage naturally benefits from buffering because DNA writes are slow compared with electronic storage. Our design uses this buffer not only to absorb writes, but also to sample the early behavior of newly written data. When a block enters the write buffer, the system initializes an activity counter for that block. Reads served from the buffer increment this counter. If the counter exceeds a configurable activity threshold before the block is flushed, the block is classified as active. Otherwise, it is classified as inactive.

DNA storage naturally benefits from buffering because DNA writes are slow compared with electronic storage. Our design uses this buffer not only to absorb writes, but also to sample the early behavior of newly written data. When a block enters the write buffer, the system initializes an activity counter for that block. Reads served from the buffer increment this counter. A block is classified as active once its buffered read count is greater than or equal to the configurable activity threshold. If the block reaches the head of the flush queue before reaching this threshold, it is classified as inactive.

The buffer flushes blocks in FIFO order. The allocator does not reorder buffered blocks based on activity because doing so would change the observation period for different blocks and bias the activity signal. Instead, the buffer acts as a fixed sampling period: each block is observed while it waits for its turn to be written. If a logical block is overwritten while it is still in the buffer, the older buffered version is removed and is never written to DNA. The new version is inserted into the buffer with a reset activity counter.

\subsection{Temperature-Aware Placement}

When a block reaches the head of the write buffer, HORIZON selects a primer pair using the block’s activity class, primer-pair temperature, and physical occupancy. Occupancy counts both live and stale strands, since all physically stored strands contribute to future read amplification.

For inactive data, the allocator prioritizes primer pairs in the following order: (1) lowest temperature, (2) fewest physically stored strands, and (3) lowest primer-pair identifier. This policy keeps inactive data away from recently active primer pairs while preserving a mostly horizontal distribution. Using occupancy as the second priority prevents inactive data from being concentrated in only a few cold primer pairs.

For active data, the allocator prioritizes primer pairs in the following order: (1) fewest physically stored strands, (2) highest temperature, and (3) lowest primer-pair identifier. The first priority minimizes immediate read amplification by placing active blocks in low-occupancy primer pairs. Temperature is used as the second priority so that, among equally occupied primer pairs, active data is placed near recently active data. As the selected primer pair accumulates more strands and becomes hotter, it becomes less likely to remain the best choice for future active blocks. This feedback prevents any single primer pair from absorbing all active data while still exploiting locality when beneficial.

\section{Evaluation}
\label{sec:evaluation}

\subsection{Experimental Setup}
We compare HORIZON with LiqSD~\cite{zhou2025liquid} in the primary evaluation and use HORIZON-Horizontal as an ablation baseline in Section~\ref{sec:placement}. The simulator records the number of strands retrieved for each read request, which allows us to measure read amplification without performing wet-lab experiments. Table~\ref{tab:sim_params} summarizes the default DNA media and block-device configuration. Unless otherwise stated, sensitivity experiments vary one HORIZON parameter at a time while keeping the remaining parameters fixed at their default values. The default configuration uses a temperature window of $10^6$ operations, a write buffer equal to 0.1\% of the device capacity, and an active/inactive threshold of 10 buffered reads. Thus, a buffered block is classified as active once it receives at least 10 reads before being flushed.
\begin{table}[h]
    \centering
    \caption{Default simulator and block-device parameters.}
    \label{tab:sim_params}
    \scriptsize
    \setlength{\tabcolsep}{1.5pt}
    
    \begin{minipage}[t]{0.48\columnwidth}
        \centering
        \textbf{DNA media}\\
        \begin{tabularx}{.8\linewidth}{@{}Xr@{}}
            \toprule
            \textbf{Parameter} & \textbf{Value} \\
            \midrule
            Strand length & 296 nt \\
            Primer length & 20 nt \\
            Payload length & 246 nt \\
            % Index length & 10 nt \\
            Strands/pair & 1,000,000 \\
            Pairs/tube & 2,000 \\
            Tubes/chip & 24 \\
            Chip count & 10 \\
            \bottomrule
        \end{tabularx}
    \end{minipage}
    \hspace{0.01\columnwidth}
    \begin{minipage}[t]{0.48\columnwidth}
        \centering
        \textbf{Block device}\\
        \begin{tabularx}{.8\linewidth}{@{}Xr@{}}
            \toprule
            \textbf{Parameter} & \textbf{Value} \\
            \midrule
            Block size & 4,096 B \\
            Strands/block & 161 \\
            Blocks/pair & 6,211 \\
            Write buffer & 0.1\% \\
            Temp. window & 1E6 \\
            Active threshold & 10 \\
             % &  \\
             % &  \\
            \bottomrule
        \end{tabularx}
    \end{minipage}

\end{table}

\begin{table}[t]
\centering
\caption{Workload trace characteristics.}
\label{tab:workloads}
\scriptsize
\setlength{\tabcolsep}{2.2pt}
\begin{tabularx}{\columnwidth}{@{}lXrrr@{}}
\toprule
\textbf{Trace} & \textbf{Description} & \textbf{Read} & \textbf{Write} & \textbf{PBA Cap.} \\
\midrule
\multicolumn{5}{@{}l}{\textit{Real-world traces}} \\
\midrule
MSR Prxy~\cite{snia-trace-block-io-388}     & Web proxy server         & 1.39 TB  & 836.0 GB & 17.8M  \\
MSR Wdev~\cite{snia-trace-block-io-388}     & Project directory server & 97.1 GB  & 9.56 GB  & 124.5M \\
MSR Hm~\cite{snia-trace-block-io-388}       & Home directory server    & 2.39 TB  & 948.8 GB & 71.7M  \\
MSR Proj~\cite{snia-trace-block-io-388}     & Project directory server & 2.08 TB  & 367.4 GB & 215.1M \\
FIU Mail~\cite{snia-trace-block-io-390}     & Mail server              & 216.5 GB & 1.68 TB  & 73.1M  \\
FIU Webusers~\cite{snia-trace-block-io-390} & Web-user server          & 4.85 GB  & 28.4 GB  & 2.08M  \\
\midrule
\multicolumn{5}{@{}l}{\textit{Synthetic filesystem traces}} \\
\midrule
F2FS Rand & F2FS random workload     & 90.6 MB  & 8.17 GB  & 16.0M \\
F2FS Seq  & F2FS sequential workload & 282.6 MB & 556.5 GB & 16.0M \\
Ext4 Rand & Ext4 random workload     & 1.42 GB  & 4.00 GB  & 5.00M \\
Ext4 Seq  & Ext4 sequential workload & 4.00 GB  & 4.00 GB  & 2.53M \\
\bottomrule
\end{tabularx}
\end{table}

We evaluate real-world block I/O traces and filesystem-based synthetic traces, summarized in Table~\ref{tab:workloads}. The real-world traces include MSR data-center traces and FIU storage traces. The synthetic traces are generated with FIO~\cite{fio} on F2FS and Ext4 and captured with blktrace~\cite{blktrace}. We include both random and sequential synthetic workloads to test whether HORIZON behaves differently under workloads with different access locality. Each trace is replayed on a fresh simulated device. The first 20\% of each trace is used as a warmup and excluded from the reported measurements.

We report read amplification (RA), defined in Equation~\ref{eq:ra}.
We compare three allocation policies: \textbf{LiqSD} (the SOTA DNA storage firmware) and \textbf{HORIZON}.

\subsection{Overall Comparison}

\begin{figure}[t]
    \centering
    %\newlength{\schemefigheight}
    %\setlength{\schemefigwidth}{0.22\textwidth}
    \begin{subfigure}[t]{0.49\textwidth}
    \centering
        %\raggedright
        \includegraphics[width=\textwidth]{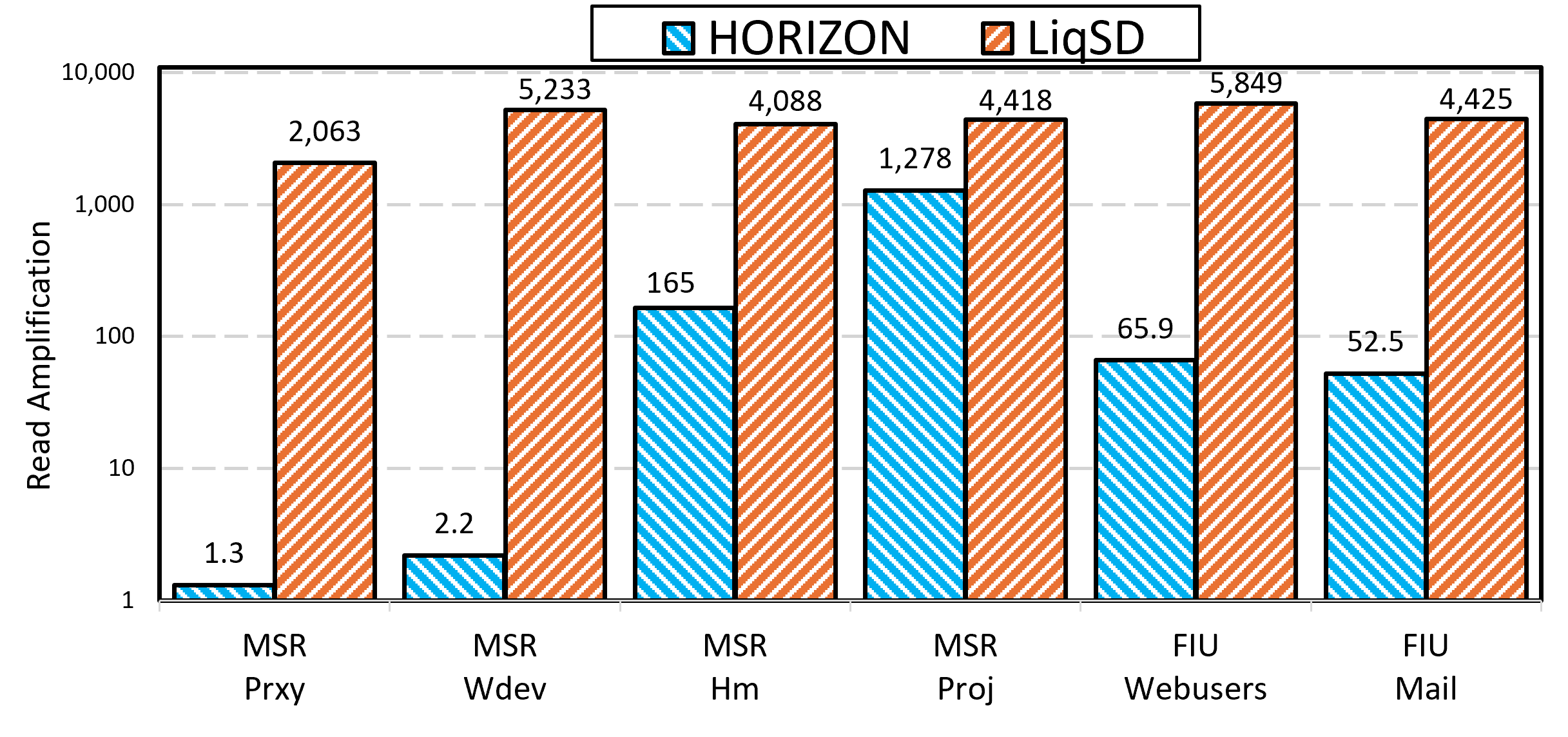}
        \caption{Trace-based workloads.}
        \label{fig:scheme_comparison_traces}
    \end{subfigure}
    % \hspace{-0.12\textwidth}
    \begin{subfigure}[t]{0.35\textwidth}
    \centering
        %\raggedright
        \includegraphics[width=0.9\textwidth]{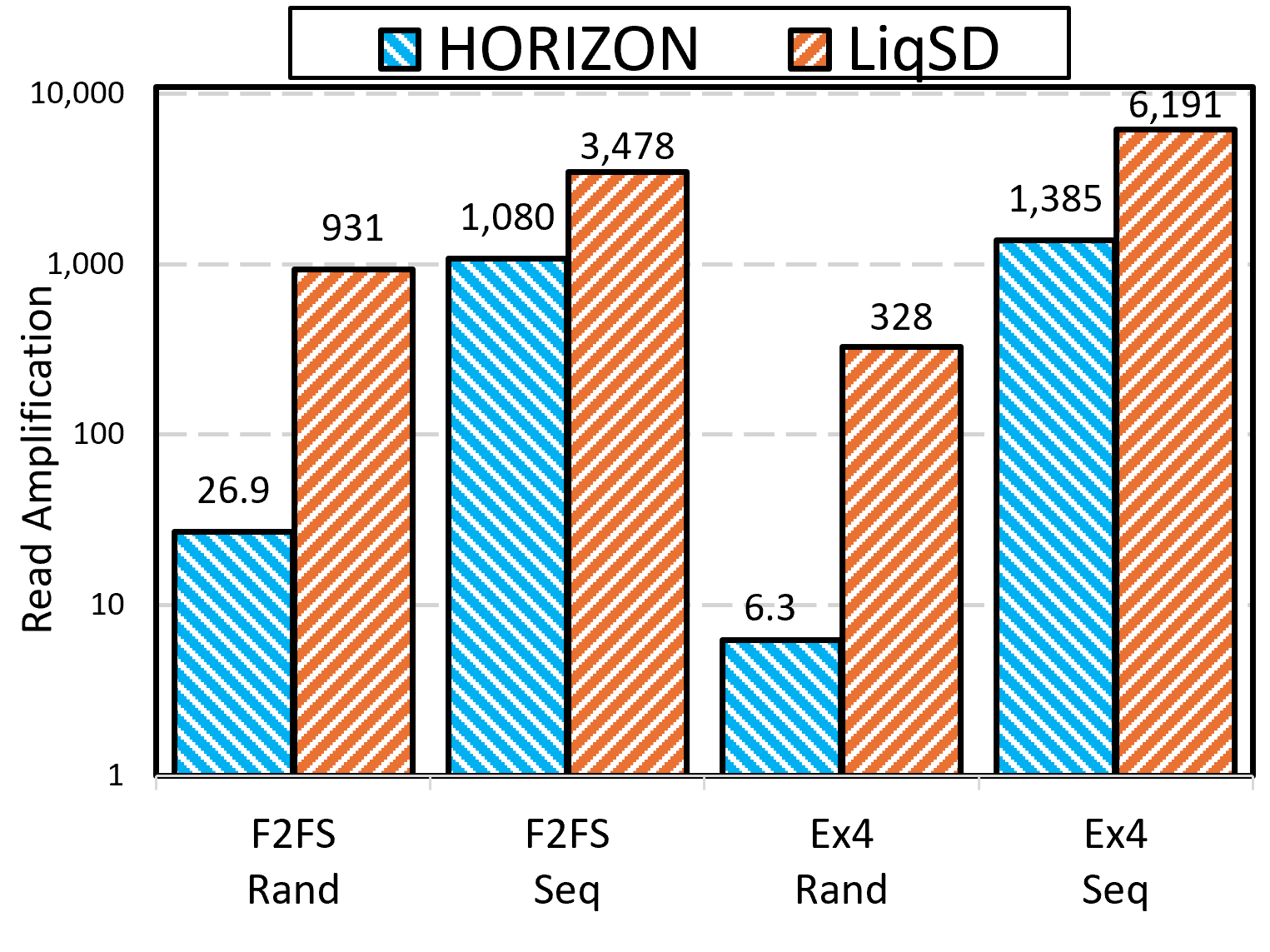}
        \caption{Synthetic filesystem workloads.}
        \label{fig:scheme_comparison_custom}
    \end{subfigure}

    % \caption{Read amplification comparison between HORIZON  and LiqSD across trace-based and custom workloads.}
    \caption{Read amplification comparison between HORIZON and LiqSD across trace-based and synthetic filesystem workloads.}
    \label{fig:scheme_comparison}
\end{figure}

%Figure~\ref{fig:scheme_comparison} compares HORIZON with LiqSD across real-world traces and synthetic filesystem workloads. HORIZON reduces read amplification on every evaluated workload, achieving a 47.8$\times$ geometric-mean reduction across all workloads, with 115.9$\times$ on real-world traces and 12.7$\times$ on synthetic filesystem workloads.
We compare HORIZON with LiqSD using real-world traces and synthetic filesystem workloads. For real-world traces in Figure~\ref{fig:scheme_comparison_traces}, on the MSR traces, HORIZON reduces RA from 2{,}063 to 1.3 on MSR Prxy and from 5{,}233 to 2.2 on MSR Wdev, corresponding to $1{,}587\times$ and $2{,}379\times$ reductions, respectively. These large gains show that HORIZON can isolate active data in low-occupancy primer pairs, preventing repeated reads from retrieving unrelated DNA. HORIZON also reduces RA by $24.8\times$ on MSR Hm, $88.8\times$ on FIU Webusers, and $84.3\times$ on FIU Mail. MSR Proj shows a smaller but still meaningful $3.46\times$ reduction, suggesting weaker locality or less effective early activity classification for this workload.

Figure~\ref{fig:scheme_comparison_custom} shows that HORIZON also improves synthetic filesystem workloads, with larger gains on random than sequential access patterns. HORIZON reduces RA by $34.6\times$ on F2FS Rand and $52.1\times$ on Ext4 Rand, compared with $3.22\times$ on F2FS Seq and $4.47\times$ on Ext4 Seq. Random workloads provide stronger activity signals through repeated accesses to subsets of the address space, allowing HORIZON to better separate active and inactive blocks. In contrast, sequential workloads provide fewer repeated buffered reads, so HORIZON behaves closer to a horizontal allocator.

\subsection{Sensitivity to Temperature Window Size}

\begin{figure}
    \centering
    \includegraphics[width=0.8\linewidth]{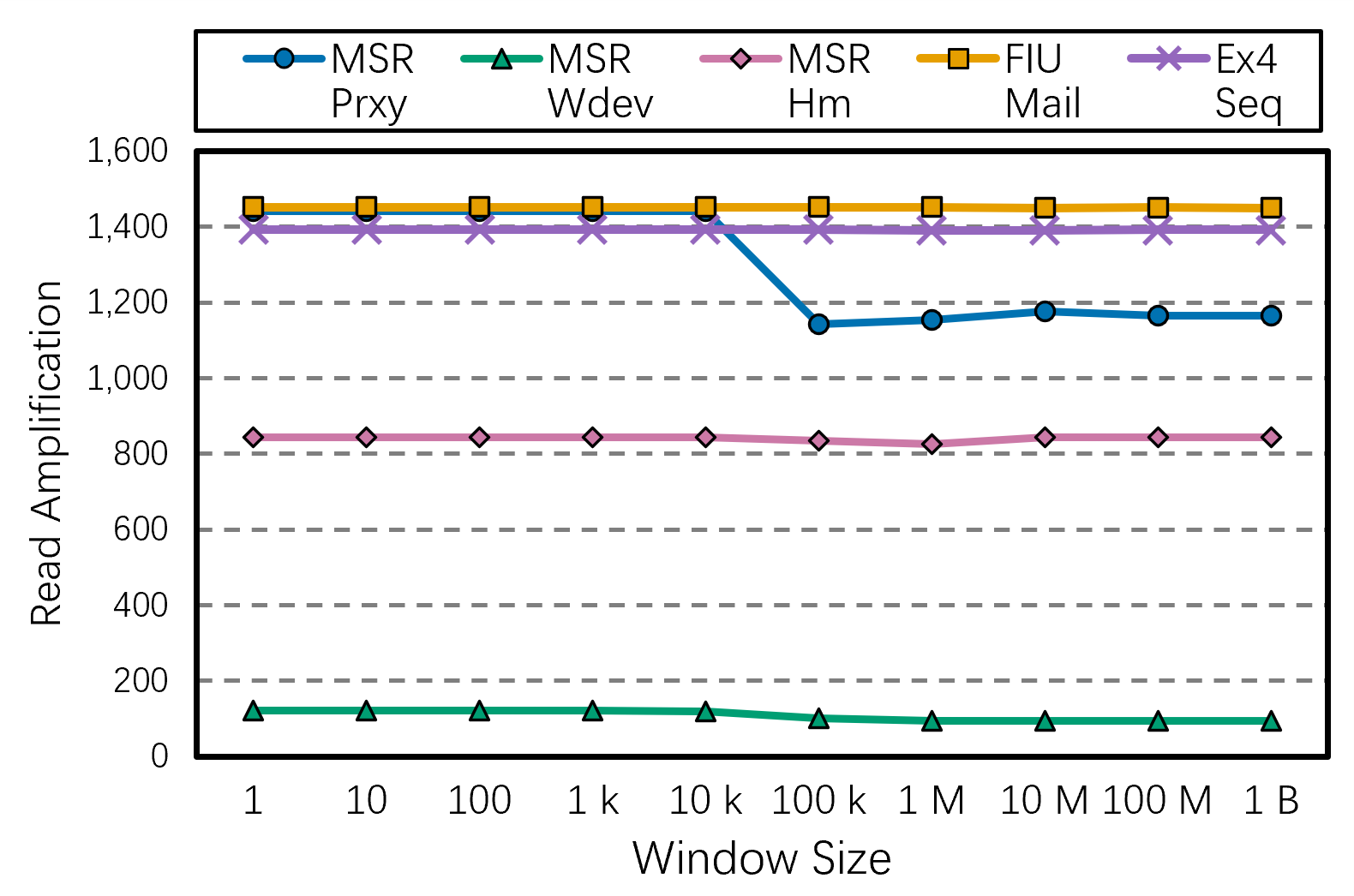}
    \caption{Impact of temperature-window size on read amplification.}
    \label{fig:window_size}
\end{figure}

Figure~\ref{fig:window_size} evaluates the effect of the temperature-window size on read amplification by isolating HORIZON’s primer-pair temperature mechanism. In this experiment, we disable the write buffer and flush writes directly to DNA, so buffered-read activity classification does not affect placement. %Therefore, the absolute RA values are not directly comparable to the full HORIZON results in Figure~\ref{fig:scheme_comparison}.
The temperature window determines how much recent history HORIZON uses to classify primer pairs as hot or cold. Small windows react quickly but can be dominated by short bursts, while large windows provide a more stable signal but adapt more slowly to workload shifts. The results show that most workloads remain stable across a broad range of window sizes, indicating that the temperature mechanism does not require precise tuning. MSR Prxy is more sensitive because it needs sufficient history to identify recently active primer pairs; once the window captures this reuse, RA decreases and remains mostly stable. We therefore use a default window of $10^6$ operations, which falls within the stable range while still allowing temperature to decay over time.

\subsection{Sensitivity to Write-Buffer}

\begin{figure}[t]
    \centering
    \begin{subfigure}[t]{0.48\linewidth}
        \centering
        \includegraphics[height=0.14\textheight,keepaspectratio]{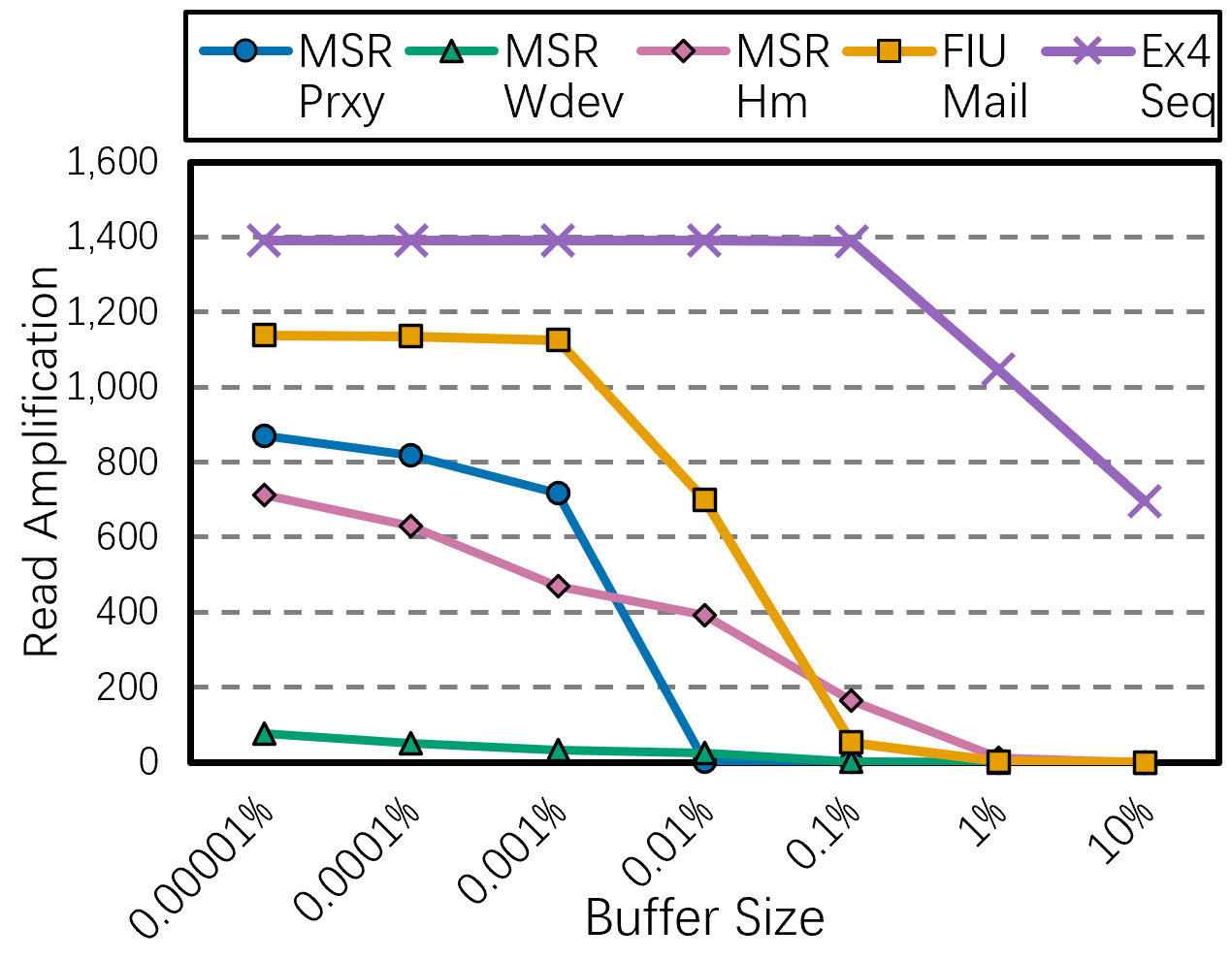}
        \caption{Buffer size.}
        \label{fig:buffer_size}
    \end{subfigure}
    % \hfill
    \begin{subfigure}[t]{0.48\linewidth}
        \centering
        \includegraphics[height=0.14\textheight,keepaspectratio]{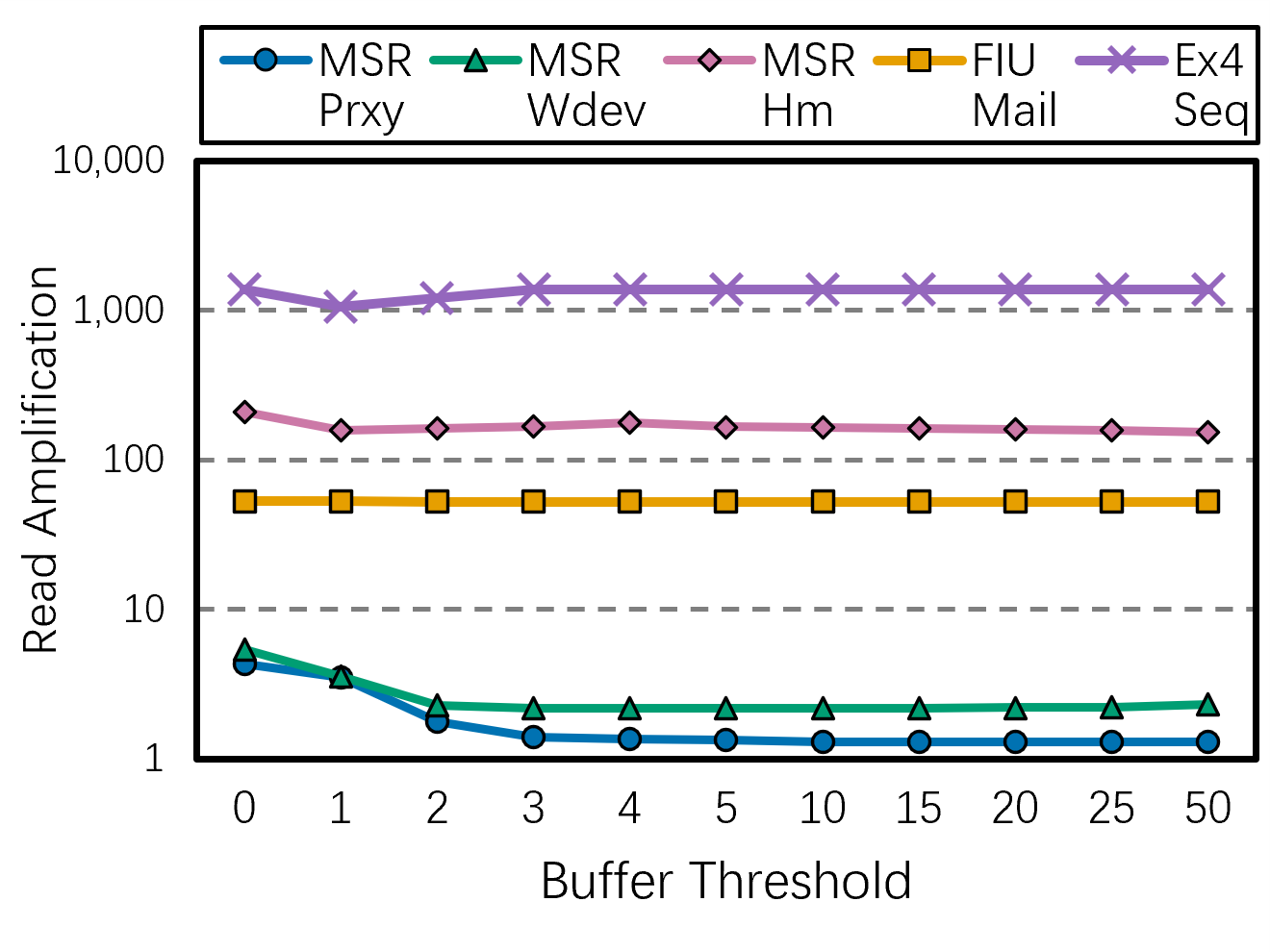}
        \caption{Buffer threshold.}
        \label{fig:buffer_threshold}
    \end{subfigure}
    \caption{Impact of buffer size and buffer threshold on read amplification.}
    \label{fig:buffer_results}
\end{figure}
%Figure~\ref{fig:buffer_size} shows that increasing the write-buffer size generally reduces read amplification. In HORIZON, the buffer both absorbs overwrites before they create stale DNA strands and provides an observation window for identifying active blocks before placement. Small buffers flush blocks too quickly, limiting activity classification, while larger buffers capture more early reads and enable better placement decisions, especially for workloads with strong short-term reuse. The benefit eventually levels off once the buffer captures most useful early activity. We therefore use 0.1\% as the default buffer size, which provides an effective observation window without requiring an unrealistically large electronic buffer.
Figure~\ref{fig:buffer_size} shows that larger write buffers generally reduce read amplification by giving HORIZON more time to absorb overwrites and observe early reads before placement. Small buffers flush blocks too quickly for accurate activity classification, while larger buffers improve placement decisions, especially for workloads with short-term reuse. The benefit levels off once the buffer captures most useful early activity. We therefore use 0.1\% as the default buffer size, which provides an effective observation window without requiring an unrealistically large electronic buffer.

Figure~\ref{fig:buffer_threshold} shows the effect of the active/inactive classification threshold. Low thresholds can misclassify incidental reads as active, while high thresholds may miss genuinely active blocks. HORIZON remains stable across a broad range of thresholds for most workloads, with MSR Prxy being the most sensitive because very low thresholds overreact to short bursts. %We therefore use a default threshold of 10, which requires repeated buffered reads while still identifying frequently reused blocks.

\subsection{Impact of Activity-Aware Placement}\label{sec:placement}

%In this section, we investigate the impact of activity-aware placement. \textbf{HORIZON-Horizontal} refers to the same write buffer but disables primer-pair temperature tracking and buffered-read activity classification. 

\begin{figure}[h]
    \centering
    \includegraphics[width=0.9\linewidth]{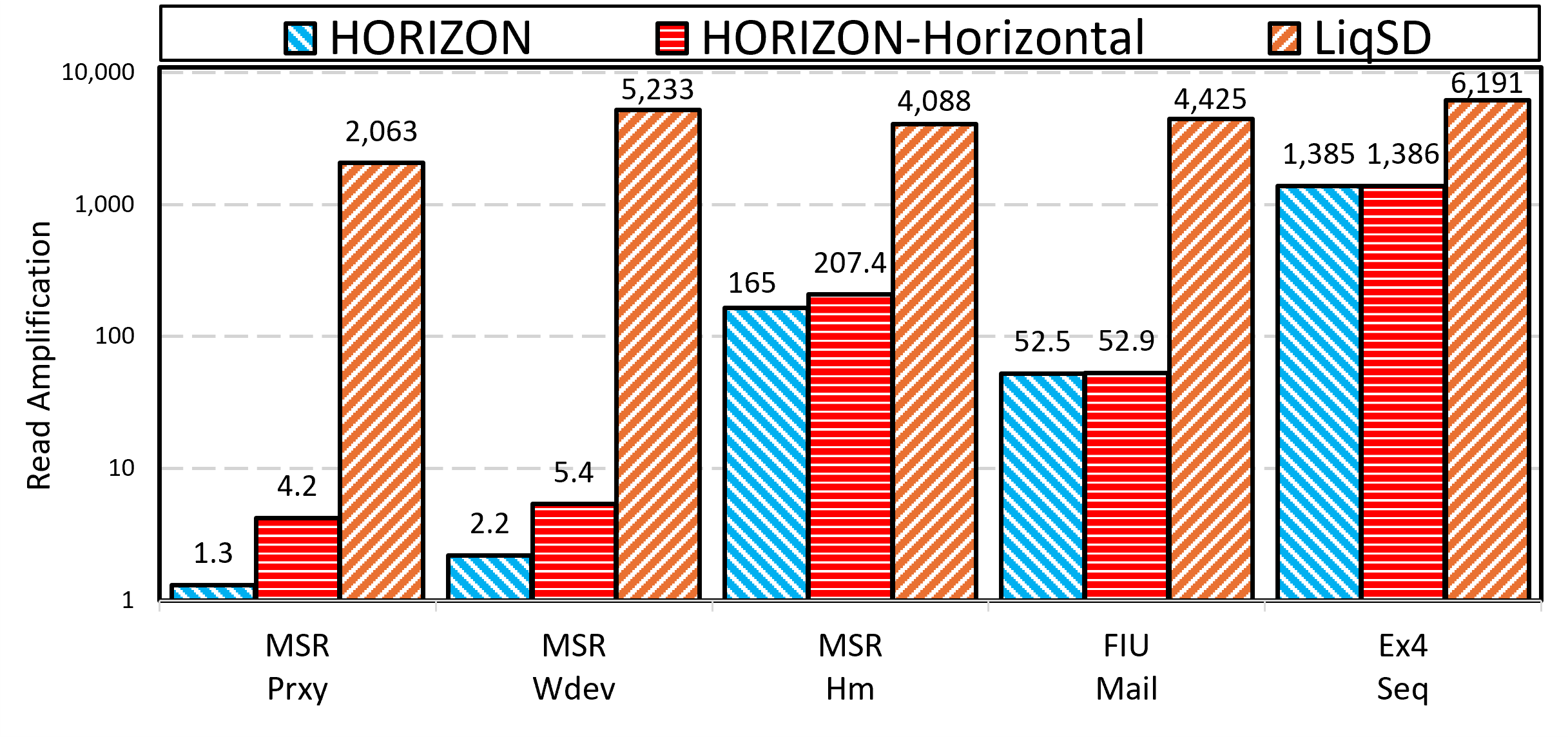}
    \caption{Ablation of HORIZON's activity-aware placement.}
    \label{fig:ablation}
\end{figure}

Figure~\ref{fig:ablation} isolates the benefit of HORIZON’s activity-aware placement beyond horizontal distribution alone. We compare HORIZON with HORIZON-Horizontal, an ablation that keeps the same write buffer and FIFO flush path but disables buffered-read activity classification and temperature-aware primer-pair selection. %Thus, HORIZON-Horizontal still spreads flushed blocks across primer pairs, but it does not distinguish active from inactive blocks or avoid placing inactive data in recently active retrieval units.

The results show that horizontal placement alone reduces read amplification compared with LiqSD, confirming that spreading blocks across primer pairs is an important foundation. However, HORIZON further reduces RA on workloads with stronger reuse, such as MSR Prxy and MSR Wdev, because it identifies active blocks before they are written to DNA and avoids polluting recently active primer pairs with inactive data. MSR Hm shows a smaller improvement, suggesting that horizontal placement provides most of the benefit, while activity and temperature tracking add refinement.

For FIU Mail and Ext4 Seq, HORIZON performs similarly to HORIZON-Horizontal because these workloads provide fewer useful activity signals, such as repeated buffered reads or concentrated reuse. In these cases, HORIZON naturally falls back toward its horizontal-placement baseline without significantly hurting performance. Overall, Figure~\ref{fig:ablation} shows that HORIZON improves read efficiency through two layers: horizontal layout reduces primer-pair occupancy, while activity-aware placement further reduces read amplification when workloads expose reusable access patterns.

\subsection{Discussion and Future Directions}
\label{sec:discussion}
HORIZON targets the same online-accessible DNA block-device model as LiqSD~\cite{zhou2025liquid}, making it most applicable to nearline archival data that receives repeated small requests. The reported RA reductions should therefore be interpreted within HORIZON's online block-storage model and not as direct reductions in latency or cost. Furthermore, the benefit of horizontal placement may decrease as primer pairs approach full capacity. Future work should consider archival workloads and jointly optimize the number of primer-pair retrievals, associated PCR operations, and usable capacity rather than focusing on strand-level RA alone.

\section{Conclusion}
\label{sec:conclusion}
This paper presented HORIZON, a read-efficient allocation policy for DNA block devices that reduces read amplification by addressing the mismatch between logical block reads and primer-pair retrieval granularity. HORIZON combines horizontal placement across primer pairs with buffered-read activity classification, primer-pair temperature tracking, and occupancy-aware allocation. Evaluation shows that HORIZON consistently outperforms LiqSD’s sequential allocation policy, achieving a 47.8$\times$ geometric-mean reduction in read amplification across all workloads.

\section*{Acknowledgment}
OpenAI ChatGPT~\cite{openai2026chatgpt} was used as a general-purpose writing assistant to revise prose in Sections~\ref{sec:introduction}-~\ref{sec:conclusion}. The authors verified all technical claims, citations, and results and take full responsibility for the paper.

\bibliographystyle{IEEEtranBST2/IEEEtran.bst}
\bibliography{bib.bib}

\end{document}

%% file: intro.tex
\section{Introduction}
\label{sec:introduction}
Digital data generation continues to grow rapidly, projected to reach up to 291 zettabytes (ZB) by 2029, driven by cloud services, artificial intelligence, etc.~\cite{idc2023,hilbert2011world,reinsel2018digitization,cisco2020air}. This growth increases the demand for long-term archival storage and places mounting pressure on conventional storage media, which face limitations in density, durability, maintenance cost, and energy consumption. DNA storage has emerged as a promising alternative because of DNA’s high molecular density and long-term chemical stability~\cite{appuswamy2019oligoarchive, ceze2019molecular,church2012next, grass2015robust, wei2024encoding}. %These properties make DNA a compelling candidate for future large-scale archival systems, but efficient data placement remains a key system-level challenge.

DNA storage encodes data into synthetic DNA strands, each containing payload data and primer sequences attached to both ends~\cite{church2012next,grass2015robust,ceze2019molecular}. A forward primer and a reverse primer form a primer pair, which acts as a logical address for a group of strands, or a retrieval unit~\cite{organick2018random}. Multiple strands can share the same primer pair, while an internal index field distinguishes individual pages within the group. During retrieval, polymerase chain reaction (PCR)~\cite{schochetman1988polymerase} selectively amplifies strands associated with the requested primer pair, avoiding the need to sequence the entire DNA pool~\cite{organick2018random,zhou2025liquid}.

To make DNA storage compatible with existing systems, recent work has proposed DNA block-device firmware that maps logical blocks to physical DNA storage locations and exposes a conventional block-storage interface over DNA media~\cite{sensintaffar2025advancing}. The state-of-the-art DNA block device, LiqSD~\cite{zhou2025liquid}, allocates blocks sequentially by filling one primer pair before moving to the next. While simple, this layout concentrates many blocks within the same retrieval unit and can co-locate frequently and rarely accessed data. As a result, each logical read may require amplifying and sequencing many unrelated strands associated with the selected primer pair, leading to high read amplification and retrieval overhead as the device fills~\cite{organick2018random,sensintaffar2025advancing}.

In this paper, we present HORIZON, a read-efficient firmware allocation policy for DNA block-device storage. The primary idea behind HORIZON is a horizontal layout that distributes writes across primer pairs, or retrieval units, rather than filling each primer pair sequentially. By spreading newly written blocks across primer pairs in a round-robin manner, HORIZON avoids rapidly concentrating large amounts of data within a single retrieval unit and reduces the amount of unrelated data retrieved during future reads. Building on this horizontal layout, HORIZON further improves allocation using activity-aware placement. It classifies newly written blocks in the write buffer as active or inactive, tracks recent primer-pair accesses using a sliding-window temperature model, and allocates blocks based on both block activity and primer-pair occupancy. Active blocks are directed to low-occupancy primer pairs to reduce immediate read amplification, while inactive blocks are placed in colder primer pairs to avoid polluting recently active retrieval units. %Together, these mechanisms reduce read amplification by controlling both how data is distributed across retrieval units and where active or inactive blocks are placed.

The rest of the paper is organized as follows. Section~\ref{sec:background} provides background on DNA storage. Section~\ref{sec:motivation} states the motivation of DNA storage and this paper. Section~\ref{sec:design} introduces the design of our firmware. Section~\ref{sec:evaluation} evaluates the schemes across different trace patterns and allocation configurations. Section~\ref{sec:conclusion} concludes this paper.

%% file: background.tex
\section{Background}
\label{sec:background}

% In DNA data storage\cite{church2012next, goldman2013towards}, digital information is encoded into nucleotide sequences and synthesized into DNA strands for physical storage. Figure~\ref{fig:dna_storage_workflow} illustrates the overall workflow. For random access, primer pairs act as addresses: writes create individual strands, while reads retrieve all strands sharing the selected primer pair. This asymmetry makes primer-pair allocation central to read amplification.

\subsection{DNA Storage Basics}
\begin{figure}[!t]
    \centering
    \includegraphics[width=\linewidth]{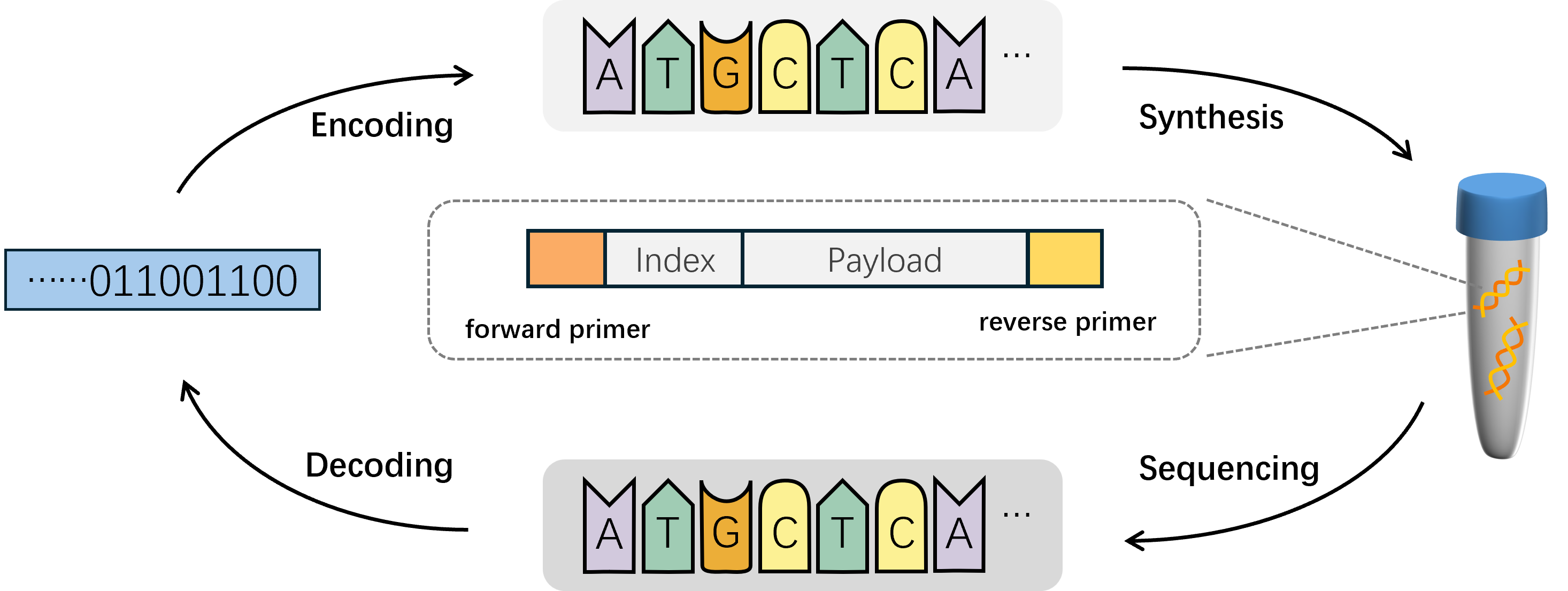}
    \caption{The overall workflow of DNA Storage.}
    \label{fig:dna_storage_workflow}
\end{figure}

Figure~\ref{fig:dna_storage_workflow} illustrates the four main phases of DNA data storage. First, digital data is encoded into nucleotide sequences~\cite{zhang2017semi, goldman2013towards, li2022hl, choi2019high}. Second, these sequences are chemically synthesized into physical DNA strands and stored in DNA tubes. Third, during readout, the system uses Polymerase Chain Reaction (PCR) to selectively amplify strands matching a chosen primer pair before sequencing the amplified DNA. Finally, the resulting nucleotide reads are decoded back into digital data.

% In DNA data storage~\cite{church2012next, goldman2013towards}, digital data is first encoded into nucleotide sequences and chemically synthesized into physical DNA strands. The synthesized strands are stored in a DNA tube until the data is later retrieved. During readout, the system selects a subset of strands using Polymerase Chain Reaction (PCR), which amplifies strands matching a chosen primer pair. The amplified DNA is then sequenced, producing nucleotide reads that are decoded back into digital data. 

% On the write path, digital data are encoded into DNA sequences and chemically synthesized into strands\cite{church2012next, bornholt2016dna}. Large objects are partitioned across multiple strands and organized via the index segment.

% Random access in DNA storage relies on Polymerase Chain Reaction (PCR)\cite{schochetman1988polymerase}. PCR selectively amplifies all DNA strands associated with a chosen primer pair, using the primer sequences as binding sites that define the amplification boundaries. Because many strands may share that primer pair, a read request for one logical block retrieves the entire primer-pair group---not just the target strand---making the primer pair the minimum unit of physical access and the root of read amplification.

% After selective amplification, the amplified strands are sequenced and decoded back into digital data. The index segment disambiguates individual payloads within the amplified group.

%\subsection{DNA Strand Layout}

A physical DNA strand used for data storage typically contains a forward primer, an index, an encoded payload, and a reverse primer. The forward and reverse primers form a primer pair, which serves as the biochemical selection address that PCR uses during readout. A primer pair is not unique to one strand; instead, many strands can share the same primer pair and are distinguished after sequencing using their index fields. During readout, PCR enriches the strands associated with the selected primer pair. Following LiqSD's system model, a read of that primer pair is charged for the entire primer-pair group, even when only a subset of its strands contains the requested data.

% During readout, PCR amplifies the entire group of strands associated with the selected primer pair, so all strands in the primer-pair group are sequenced even when only a subset contains the requested data.

% Therefore, a read to one logical block may amplify and sequence other strands stored under the same primer-pair group, making primer-pair organization a key source of read amplification.

% A physical DNA strand designed for data storage typically comprises \textbf{primers} (short sequences at both ends), an \textbf{index}, a \textbf{payload} carrying the encoded digital data, and encoding redundancy for synthesis and sequencing errors. The primer sequences on each strand are binding sites for enzymatic amplification; a \textbf{primer pair}---the forward and reverse primers used together---serves as the logical address for selecting a group of strands during readout, not as a label unique to a single strand. Because many strands, pages, or payload fragments can share the same primer pair, the \textbf{index} allows payloads associated with that pair to be distinguished after readout. This many-to-one mapping between strands and primer pairs is why read operations amplify non-target as well as target data.

DNA storage systems are constrained by both strand length and primer availability. DNA storage systems commonly use strands shorter than 300 nucleotides, which limits the amount of payload data that can be stored in each physical strand~\cite{church2012next, grass2015robust, organick2018random, zhou2025liquid}. Primer libraries are also limited in practice to approximately 14,000 usable primer pairs~\cite{organick2018random, song2021multidimensional,yamamoto2008large}. Together, these constraints require many strands to share each primer pair, making primer-pair grouping a central storage-layout decision.

% Due to biochemical and hardware constraints, individual DNA strand lengths are strictly limited; mainstream DNA storage research typically uses strands of fewer than 300 nucleotides~\cite{church2012next, grass2015robust, organick2018random, zhou2025liquid}. The number of usable primer pairs is also limited in practice; in a standard random-access library, prior work demonstrated approximately 14{,}000 usable primer pairs (roughly 28{,}000 individual primer sequences)~\cite{organick2018random}.

\subsection{DNA Storage Block Device}

A block device exposes storage as fixed-size addressable units, such as 4 KiB blocks. LiqSD~\cite{zhou2025liquid} introduced a DNA block-device architecture that maps logical block requests to physical DNA storage locations. In this model, DNA tubes store archival payload strands, while electronic storage maintains metadata, buffers writes, and serves cached read data. This hybrid organization allows the system to present a conventional block interface while using DNA as the persistent storage medium.

% A block device is a storage abstraction that organizes and randomly accesses data in fixed-size units, such as 4 KiB blocks. The LiqSD\cite{zhou2025liquid} system first introduced the design concept for a holistic DNA block device. In this model, the physical architecture is a hybrid of DNA and electronic storage: DNA tubes hold archival payload strands, while electronic storage maintains metadata, buffers writes, and serves as a low-latency cache.

\begin{figure}[t]
    \centering
    \includegraphics[width=0.8\linewidth]{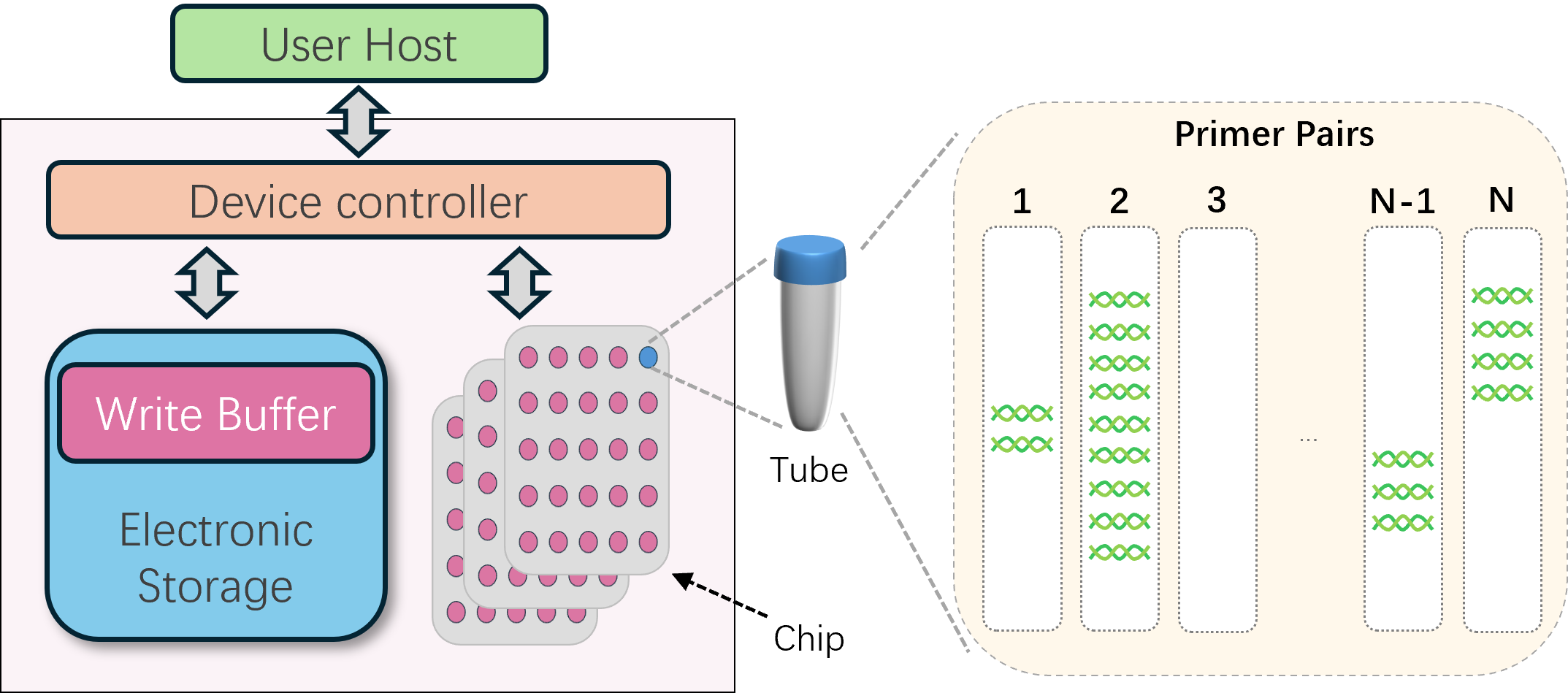}
    \caption{The overall architecture of DNA Block Device.}
    \label{fig:dna_block_device}
\end{figure}

Figure~\ref{fig:dna_block_device} shows the DNA block-device architecture assumed in this work. The device controller translates block requests into DNA storage operations and manages logical-to-physical address mapping. Electronic storage maintains persistent metadata, buffers writes, and caches recently read data. The chip organizes multiple DNA tubes and coordinates molecular operations across them. Each tube is a physical DNA pool containing strands selected by an associated primer library; it acts as non-volatile archival storage and as the minimum erase unit, since erasing requires destroying all strands within a tube.

\textbf{Read Amplification Model:} Operation granularity is asymmetric in random access-based DNA block devices: the DNA strand is the minimum write unit, and the primer-pair group is the minimum read unit. Each primer pair identifies a group of strands. Reading a logical block from DNA requires retrieving the primer pair containing that block. The requested block may require only a small subset of the strands assigned to that primer pair, but the biochemical retrieval process amplifies and sequences the primer-pair group as a whole. We therefore define read amplification as the ratio between the physical DNA retrieved and the DNA required by the logical request:

\begin{equation}
RA =
\frac{\text{DNA strands retrieved for the logical read request}}
{\text{DNA strands required for the requested block}}
\label{eq:ra}
\end{equation}

For reads that access DNA, the numerator includes all strands physically stored under the selected primer pair, including stale or invalidated versions. %This is because DNA updates are append-only at the strand level: rewriting a logical block writes a new physical version and updates metadata to point to the new location, but the old DNA strands remain present. Removing stale strands would require erasing the physical DNA tube, which is expensive because it destroys all strands in the tube. For reads served entirely from the write buffer, no DNA strands are retrieved, so the numerator is zero, and the read contributes an RA value of zero. As a result, the reported average RA captures both the cost of DNA retrievals and the benefit of avoiding DNA access through buffered reads.

%% file: motivation.tex
\section{Extended Motivation}
\label{sec:motivation}

\begin{figure}[htbp]
    \centering
    
    \begin{subfigure}[b]{0.48\linewidth}  
        \centering
        \includegraphics[width=\linewidth]{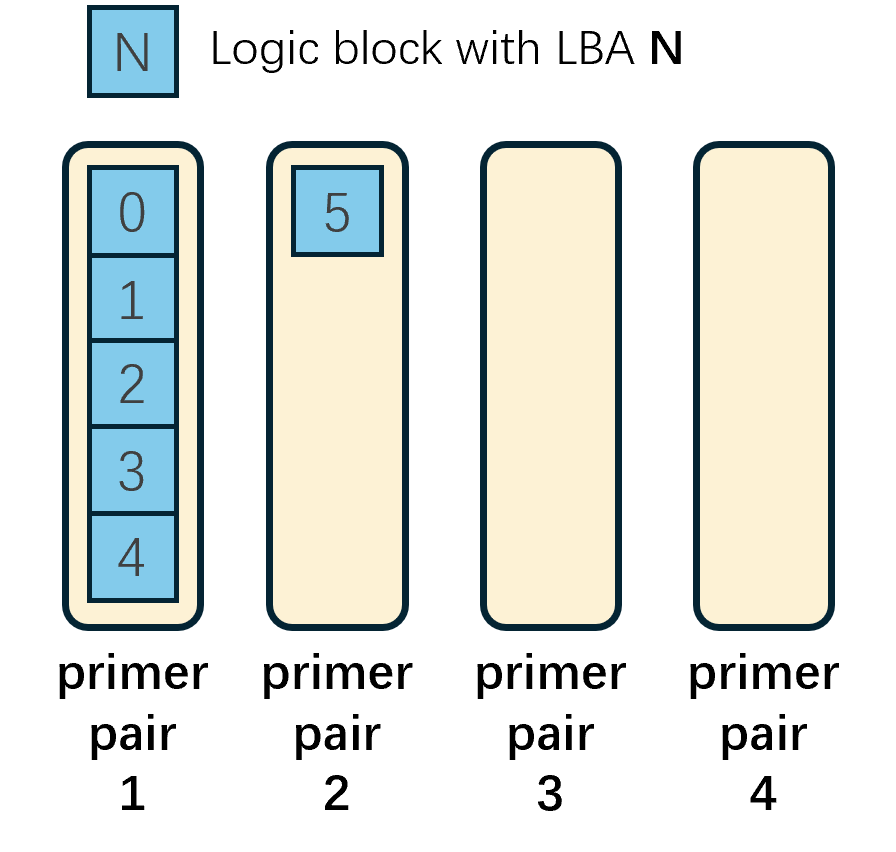} 
        \caption{Vertical distribution}         
        \label{fig:vertical_dis}
    \end{subfigure}
    \hfill
    \begin{subfigure}[b]{0.48\linewidth}  
        \centering
        \includegraphics[width=\linewidth]{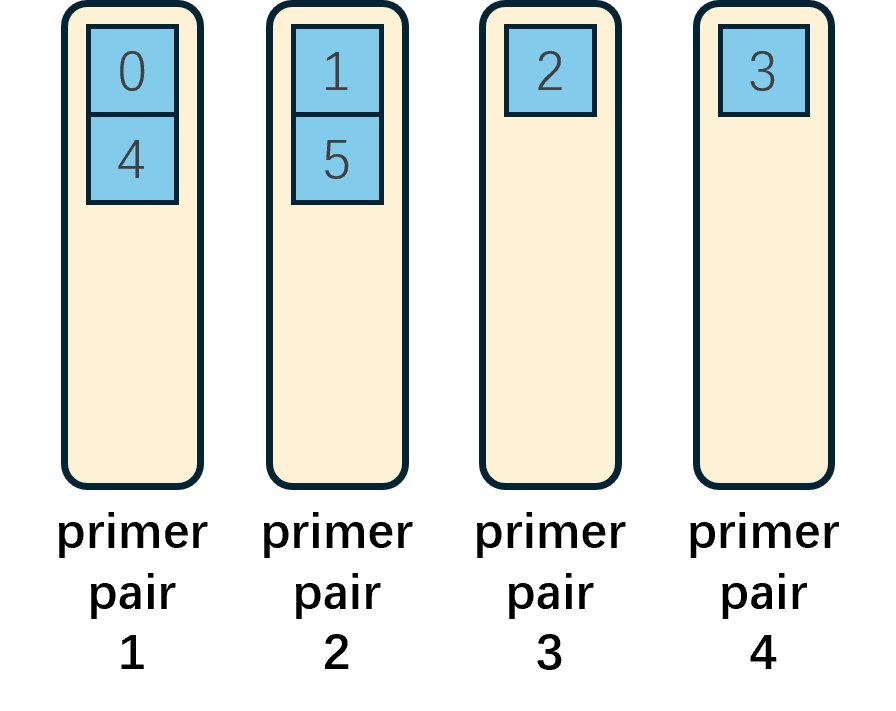}
        \caption{Horizontal distribution}         
        \label{fig:horizontal_dis} 
    \end{subfigure}
    
    \caption{Comparison of vertical and horizontal physical data distributions across primer pairs.}
    \label{fig:main_results} 
\end{figure}

Unlike conventional block devices, the DNA block-device model retrieves data at primer-pair granularity, so one logical block read can return many strands unrelated to the requested block. Following LiqSD~\cite{zhou2025liquid}, we use read amplification as a relative system-level metric of this excess DNA retrieval rather than as a direct model of wall-clock latency or monetary cost. Primer-pair allocation therefore affects how much unrelated DNA is retrieved per logical read as the device fills.
% Unlike conventional block devices, each DNA read requires PCR amplification and sequencing, so read amplification directly increases retrieval latency and cost. Primer pair allocation therefore affects how expensive future reads become as the system fills.

The state-of-the-art DNA block device, LiqSD~\cite{zhou2025liquid}, does not consider block activity when placing data. It maps logical blocks to primer pairs using sequential allocation: incoming blocks are appended to the same primer pair in write order until that primer pair is full. Physically, this creates a vertical distribution in which the strands of many blocks are colocated within one primer-pair group, as shown in Figure~\ref{fig:vertical_dis}. Because each read amplifies the entire primer-pair group, this layout concentrates many strands under a single address and drives up read amplification. A better physical layout is horizontal distribution: distributing blocks across more primer pairs so that each group holds fewer strands, as illustrated in Figure~\ref{fig:horizontal_dis}. 

% \color{red}
% (remove)
% \color{black}
% Round-robin allocation is a straightforward way to achieve horizontal placement by cycling through primer pairs in order. However, round-robin does not use block activity when placing data. It treats every block the same regardless of how often it will be read, ignores early read behavior observed while blocks remain in the write buffer, and does not consider which primer pairs have been accessed recently. As the system fills, frequently read blocks can therefore still be written into primer-pair groups that are already crowded or unrelated to their access pattern, so read amplification can continue to grow even when data is spread horizontally.

Reducing read amplification therefore requires placement that uses block activity, not placement order alone. An allocator should observe early read behavior while blocks remain in the write buffer, distinguish active from inactive blocks, and account for recent primer-pair access when choosing where to commit data. Active blocks should be directed toward low-occupancy primer pairs to limit immediate read amplification, while inactive blocks should be kept away from recently hot primer pairs so they do not pollute active retrieval units. Section~\ref{sec:design} presents our allocation policy in detail.